\def\dir{./}
\documentclass[useAMS,usenatbib]{\dir mn2e}
\usepackage[fleqn]{amsmath}
\usepackage{amssymb}
\usepackage[super]{nth}
\usepackage{natbib}
\usepackage{graphicx}
\usepackage{float}
\usepackage{mathrsfs}
\usepackage{mathtools}
\usepackage{multirow}
\usepackage[usenames,dvipsnames]{color}
\usepackage{\dir aas_macros}
\usepackage{comment}
\usepackage{subfigure}
\usepackage{hyperref}
\usepackage{mathrsfs}
\usepackage{epstopdf, dcolumn}
\DeclareGraphicsExtensions{.pdf}
\usepackage{wasysym}
\usepackage{dashrule}
\usepackage{xcolor}
\usepackage{threeparttable}

\newcommand{\sub}[1]{{\rm #1}}

\title{Dependence of galaxy clustering on UV-luminosity and stellar mass at $z \sim 4 - 7$}

\author[Qiu et al.]{Yisheng Qiu$^{1,2}$\thanks{E-mail: yishengq@student.unimelb.edu.au},
                    J. Stuart B. Wyithe$^{1,2}$\thanks{E-mail: swyithe@unimelb.edu.au}, 
                    Pascal A. Oesch$^{3}$, 
                    Simon J. Mutch$^{1,2}$, 
					\newauthor
                    Yuxiang Qin$^{1,2}$,
                    Ivo Labb\'e$^{4}$,
                    Rychard J. Bouwens$^{5}$,                                     
                    Mauro Stefanon$^{5}$,
                    Garth D. Illingworth$^{6}$   
	\\$^{1}$School of Physics, University of Melbourne, Parkville, VIC 3010, Australia
	\\$^{2}$ARC Centre of Excellence for All Sky Astrophysics in 3 Dimensions (ASTRO 3D)
	\\$^{3}$Department of Astronomy, University of Geneva, Chemin des Maillettes 51, 1290 Versoix, Switzerland
	\\$^{4}$Centre for Astrophysics \& Supercomputing, Swinburne University of Technology, PO Box 218, Hawthorn, VIC 3112, Australia
	\\$^{5}$Leiden Observatory, Leiden University, NL-2300 RA Leiden, Netherlands
	\\$^{6}$UCO/Lick Observatory, University of California, Santa Cruz, CA 95064, USA}
\begin{document}

\pagerange{\pageref{firstpage}--\pageref{lastpage}} \pubyear{2017}
\maketitle
\label{firstpage}

\begin{abstract}
We investigate the dependence of galaxy clustering at $z \sim 4 - 7$ on UV-luminosity and stellar mass. Our sample consists of $\sim$ 10,000 Lyman-break galaxies (LBGs) in the XDF and CANDELS fields. As part of our analysis, the $M_\star - M_\sub{UV}$ relation is estimated for the sample, which is found to have a nearly linear slope of $d\log_{10} M_\star / d M_\sub{UV} \sim 0.44$. We subsequently measure the angular correlation function and bias in different stellar mass and luminosity bins. We focus on comparing the clustering dependence on these two properties. While UV-luminosity is only related to recent starbursts of a galaxy, stellar mass reflects the integrated build-up of the whole star formation history, which should make it more tightly correlated with halo mass. Hence, the clustering segregation with stellar mass is expected to be larger than with luminosity. However, our measurements suggest that the segregation with luminosity is larger with $\simeq 90\%$ confidence (neglecting contributions from systematic errors). We compare this unexpected result with predictions from the \textsc{Meraxes} semi-analytic galaxy formation model. Interestingly, the model reproduces the observed angular correlation functions, and also suggests stronger clustering segregation with luminosity. The comparison between our observations and the model provides evidence of multiple halo occupation in the small scale clustering.

\end{abstract}


\section{Introduction}
Galaxy clustering provides a probe of the host halo mass of galaxies. The clustering strength is commonly described by the two-point correlation function, which measures the probability of finding galaxy pairs at given spatial separations. \cite{1996MNRAS.282..347M} used the extended Press-Schechter formalism \citep{1991ApJ...379..440B} to show that the ratio between the correlation functions of halos and the underlying matter depends on halo mass. This ratio is known as bias. Since galaxies reside in halos, the bias links galaxy clustering to the mass of their host halos (see \citealp{2002PhR...372....1C} for a review).
\par
The dependence of clustering strength on galaxy properties is known as clustering segregation, and reveals the correlation between galaxy properties and halo mass. At high redshifts, clustering segregation is observed for Lyman-break galaxies (LBGs) with UV-luminosity \citep{2006ApJ...642...63L,2009A&A...498..725H,2014ApJ...793...17B,2016ApJ...821..123H,2018PASJ...70S..11H} and stellar mass \citep{,2017ApJ...841....8I,2018A&A...612A..42D}. One basic conclusion from these studies is that more luminous and larger stellar mass galaxies are more clustered, and therefore reside in more massive halos.
\par
In the context of hierarchical galaxy formation, this correlation between halo and galaxy properties is unsurprising, since halo mass is closely related to the gas reservoir available for star formation and those processes which impact on it. For instance, in the low mass regime, supernova (SN) feedback can effectively suppress star formation \citep{2013MNRAS.428.2741W,2014MNRAS.445..581H,2014MNRAS.443.3435D}. Therefore, it is of particular interest to explore which galaxy property is more tightly correlated with the host halo mass. While UV-luminosity is directly related to the current star formation rate, stellar mass provides integrated information over the star formation history. For this reason, it is expected that, when splitting the same sample by luminosity and stellar mass, clustering segregation with stellar mass should be larger than with UV magnitude. 
\par
In order to observe the difference between clustering segregation with stellar mass and luminosity, it is essential that the stellar mass is measured from SED fitting including rest-frame optical photometry. Recently, \citet{2016ApJ...821..123H} carried out an analysis of clustering segregation with stellar mass in similar fields to our work. However, the stellar mass used in that study was obtained from a simple conversion of the UV- luminosity to mass using the $M_\star - M_\sub{UV}$ relation. Thus, the analysis could not self-consistently infer any difference of the clustering segregation between stellar mass and UV-luminosity. In this paper, we measure stellar masses from SED fitting including rest-frame optical Spitzer/IRAC data, which allows us to measure the clustering segregation in stellar mass at these redshifts for the first time.
\par
Semi-analytic models (SAMs) of galaxy formation are based on halo merger trees provided by N-body simulations, and evolve galaxy properties within these halos using analytic or empirical prescriptions of baryonic physics and feedback processes. Since the theory which links galaxy clustering to dark matter halos also relies on N-body simulations \citep[e.g][]{1996ApJ...462..563N,2010ApJ...724..878T}, a comparison between observed clustering and that predicted by a SAM tests the link between galaxy properties and halo mass. In this study, we compare our clustering measurements with results from the \textsc{Meraxes} SAM. This model has been shown to be successful in reproducing UV-luminosity functions and stellar mass functions over a wide range of redshifts \citep{2016MNRAS.462..250M,2016MNRAS.462..235L,2017MNRAS.472.2009Q}.
\par
This paper is organised as follows. We describe the observational catalogue used in the analysis, and measure the $M_\star - M_\sub{UV}$ relation in Section \ref{data}. Methods to measure the angular correlation function (ACF) and galaxy bias are introduced in Section \ref{methods}, and results are demonstrated in Section \ref{results}. We perform the comparison between the observations and predictions from \textsc{Meraxes} in Section \ref{compare}. Finally, the work is summarised in Section \ref{sum}. Throughout the paper, unless specified, the cosmology of $(h, \: \Omega_\sub{m}, \: \Omega_\sub{b}, \: \Omega_\Lambda, \: \sigma_8)=(0.678, \: 0.308, \: 0.0484, \: 0.692, \: 0.815)$ \citep{2016A&A...594A...1P} is assumed. Magnitudes are in the AB system.

\begin{figure*}
  \includegraphics[width=16cm]{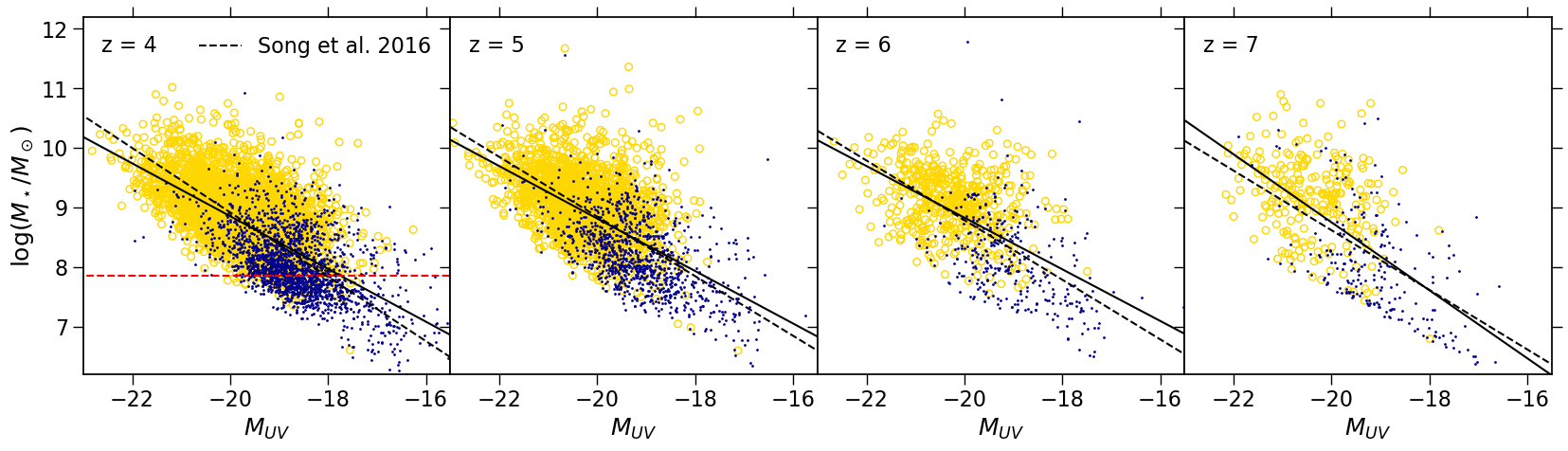} 
  \caption{Scatter plots of the $M_\star - M_\sub{UV}$ relations. The absolute magnitude $M_\sub{UV}$ is based on the flux in the filter whose centre wavelength is closest to rest-frame $1600 \text{\AA}$ and the stellar mass is measured by SED fitting including Spitzer data. For yellow empty circles, the sources have at least one Spitzer band ($3.6 \mu m$ and $4.5 \mu m$) with S/N$ > 3$, while small blue dots show the remaining sources. Black solid lines are best-fit results. The corresponding parameters are summarised in Table \ref{LTM}. For comparison, black dashed lines show the results from \citet{2016ApJ...825....5S}, which are converted to a \citet{2003PASP..115..763C} IMF by subtracting 0.24 dex in the stellar mass. The dashed horizontal line shows the lower bound of the least massive stellar mass bin for the clustering measurement at $z \sim 4 $.}
  \label{relation}
\end{figure*}

\section{Data} \label{data}
The galaxy sample used for our measurements is based on the photometric catalogue from \citet{2015ApJ...803...34B}, who selected Lyman break galaxies (LBGs) at $z\sim4-7$ based on the \textit{Hubble Space Telescope} (HST) data in all the CANDELS fields, as well as the very deep XDF and HUDF09 parallel fields. In particular, our sample is drawn from the \textit{XDF} \citep{2013ApJS..209....6I}, HUDF-091 and HUDF-092 \citep{2011ApJ...737...90B}, \textit{CANDELS-GN} and \textit{CANDELS-GS} \citep{2011ApJS..197...35G,2011ApJS..197...36K}, \textit{ERS} \citep{2011ApJS..193...27W}, and \textit{CANDELS-UDS}, \textit{CANDELS-COSMOS} and \textit{CANDELS-EGS} \citep{2011ApJS..197...35G,2011ApJS..197...36K}.
These survey regions span an aggregate of $\sim 700 \, \textrm{arcmin}^2$ in the sky, and $\sim 10,000$ LBGs are identified. Photometric redshifts of these sources are estimated using the EAZY code \citep{Brammer08}.
For more information on the LBG selection and the photometric redshifts see \citet{2015ApJ...803...34B}.

We combine the HST photometry with the large archive of \textit{Spitzer/IRAC} legacy data available in the CANDELS fields \citep{Ashby13,Ashby15}, which includes the ultra-deep IGOODS/IUDF and GREATS surveys \citep[][Labb\'e et al. 2018, in prep.]{2015ApJS..221...23L} in the GOODS fields. IRAC photometry is measured in circular apertures after subtracting the contaminating flux of neighboring galaxies using the code $mophongo$ \citep[][Labb\'e et al., 2018, in prep.]{Labbe06}, which is similar to the code TPHOT \citep{Merlin16}.
\par
We measure stellar masses of galaxies based on SED fitting to the HST+Spitzer photometry using ZEBRA+ \citep{Oesch10}. 
The synthetic template set used here is based on \citet{BC03} with a constant star-formation history, sub-solar metallicities (0.2 Z$_\odot$) and a \citet{2003PASP..115..763C} initial mass function (IMF). Nebular
continuum and emission lines are added self-consistently based on the number of ionizing photons emitted by each SED and assuming line ratios relative to H$\beta$ as tabulated by \citet{Anders03}.
Dust extinction is applied using the attenuation curve by \citet{Calzetti00}.

Following \citet{2015ApJ...803...34B}, the absolute magnitudes, $M_\sub{UV}$, are computed based on the fluxes in the photometric band that is closest to rest-frame $1600 \text{\AA}$. 
We first fit the $M_\star - M_\sub{UV}$ relation for the LBG sample, which will be used in our clustering analysis to compare stellar mass and luminosity segregation. The form of the relation is assumed to be
\begin{equation}
 \log_{10} M^\sub{Fit}_\star = \frac{d \log_{10} M_\star}{dM_\sub{UV}} \left (M_\sub{UV} + 19.5 \right ) + \log_{10} M_{\star (M_{\rm UV = -19.5})},
\end{equation}
where mass is in units of $M_\odot$. The log-likelihood is then constructed as 
\begin{equation}
\ln \mathcal{L} = -\frac{1}{2} \sum\limits_i \left [ \frac{\log_{10}({M^\sub{Obs}_{\star i}}_/M^\sub{Fit}_\star)^2}{\Delta^2} + \ln (2 \pi \Delta^2)\right],
\end{equation}
where the sum is over all LBGs, and $\Delta$ is a mass-independent free parameter representing scatter in the $M_\star - M_\sub{UV}$ relation. We adopt a Bayesian approach to perform the fit, assume constant priors for all parameters, and apply the \textsc{emcee} MCMC sampler developed by \citet{2013PASP..125..306F}. The resulting $M_\star - M_\sub{UV}$ relations are shown in Figure \ref{relation}. Best-fit parameters are given in Table \ref{LTM}. We find that the $M_\star - M_\sub{UV}$ relations are close to linear ($M_\star \propto L)$ and that the scatter in stellar mass at fixed luminosity is $\sim 0.5$ dex. Even though our best-fit slopes are slightly shallower, our measurements are consistent with the recent study from \citet{2016ApJ...825....5S}. 
\par
In this work, every galaxy in our HST sample has an estimate of stellar mass irrespective of the quality of Spitzer data. Low S/N ratios of Spitzer bands could make the stellar masses less precise. To investigate this, in Figure \ref{relation}, galaxies that have at least one Spitzer band ($3.6 \mu m$ and $4.5 \mu m$) with S/N$ > 3$ are shown as yellow empty circles in Figure \ref{relation}, while the others are shown as small blue dots. No systematic offset is found between them. Since the sample is large enough at $z \sim 4$, we use multiple stellar mass and luminosity bins for the clustering measurements, and avoid using bins that have no lower bound to reduce possible effects due to low S/N ratios of Spitzer data. The lower bound for the least massive stellar mass bin is shown as red dashed line in the corresponding panel of Figure \ref{relation}. At all other redshifts, limited by the sample size, we include all galaxies and use two bins to examine clustering segregation with stellar mass and luminosity. The fraction of LBGs that are included and have at least one Spitzer band with S/N$ > 3$ is 74\%, 63\%, 53\%, and 47\% at $z \sim 4, 5, 6$, and $7$ respectively. The advantage of this approach is that the completeness of sample LBGs, which is defined by the selection, is not affected by Spitzer data, and it is important to use the same set of galaxies to compare the clustering segregation between UV-luminosity and stellar mass. We will discuss possible effects on clustering measurements due to uncertainties in stellar mass in Section \ref{scatter}.

\begin{table}
  \caption{Best-fit parameters of the $M_\star - M_\sub{UV}$ relations}
  \label{LTM}
  \begin{tabular}{c|c|c|c}
    $\bar{z}$ & $d \log_{10} M_\star / dM_\sub{UV}$ & $\log_{10}M_{\star (M_{\rm UV = -19.5})} $ & $\Delta$ \\
    \hline
    3.8 & $-0.44^{+0.01}_{-0.01}$ & $8.64^{+0.01}_{-0.01}$  & $0.501^{+0.005}_{-0.005}$\\
    5.0 & $-0.44^{+0.01}_{-0.01}$ & $8.60^{+0.01}_{-0.01}$  & $0.555^{+0.008}_{-0.007}$\\
    5.9 & $-0.43^{+0.02}_{-0.02}$ & $8.61^{+0.02}_{-0.02}$  & $0.621^{+0.017}_{-0.013}$\\
    6.8 & $-0.57^{+0.03}_{-0.03}$ & $8.47^{+0.03}_{-0.03}$  & $0.734^{+0.028}_{-0.020}$\\
  \end{tabular}
  \begin{tablenotes}
    \item Notes. - Mass unit is $M_\odot$. Quoted errors are $16 \%$ and $84 \%$ percentiles of the marginalised distributions estimated by the MCMC sampler. 
  \end{tablenotes}

\end{table}

\begin{figure*}
  \includegraphics[width=16cm]{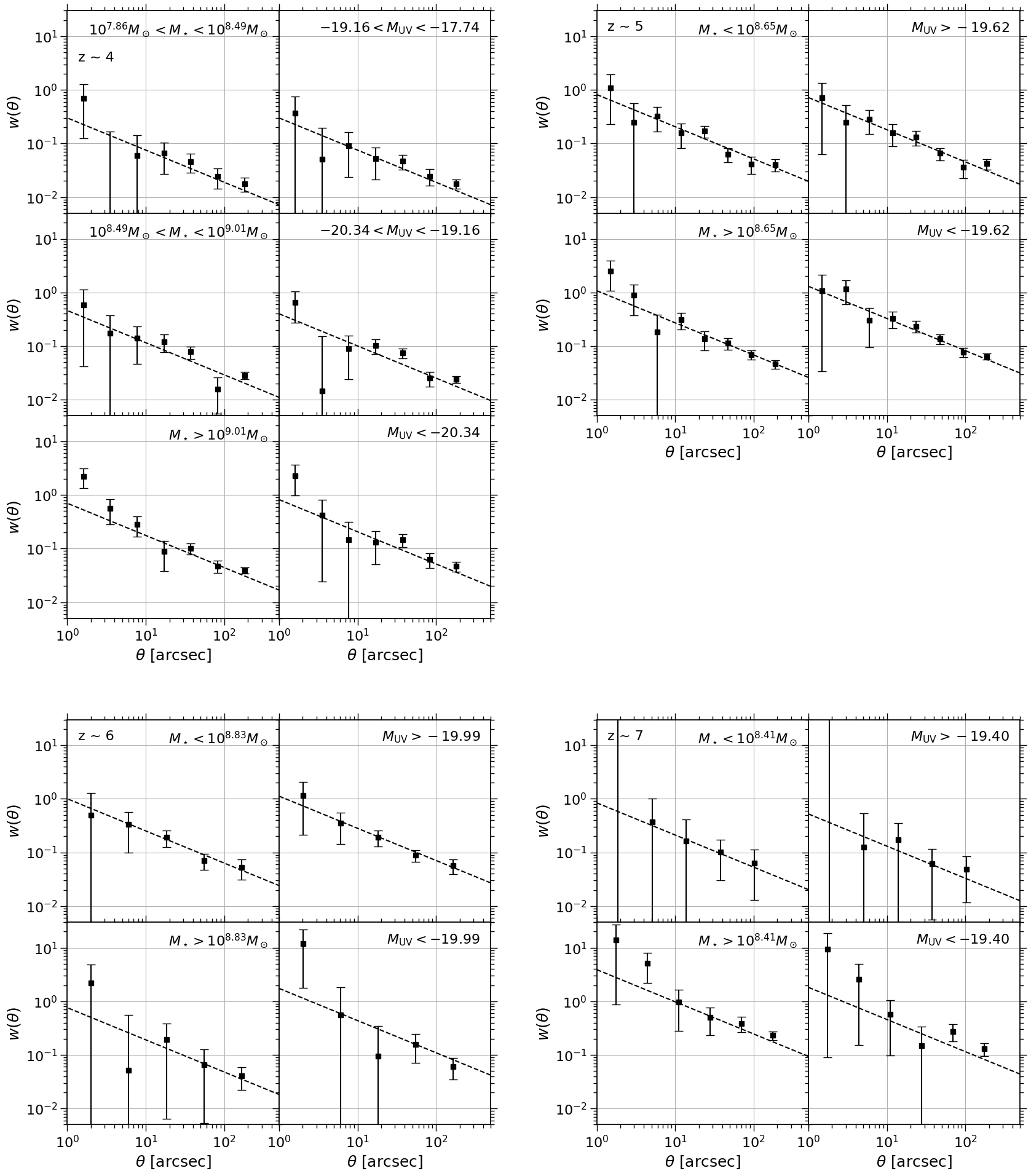} 
  \caption{Measured ACFs and their best-fit power laws $A_\omega (\theta/1'')^{-0.6}$. Top left, top right, bottom left, and bottom right panels show the measurements at $\bar{z} = 3.8$, $5.0$, $5.9$, and $6.8$ respectively. For each panel, the first and second columns illustrate the stellar mass and luminosity subsamples respectively. In all plots, black squares with error bars are measured ACFs, which are averaged over all fields using inverse-variance weighting, and dashed lines are best-fit power laws.}
  \label{acfs}
\end{figure*}

\section{Clustering} \label{methods}
\subsection{Estimating the angular correlation function} \label{sACF}
Our approach follows \citet{2014ApJ...793...17B}. We start by determining the angular correlation functions (ACFs), which measure the excess probability of finding galaxy pairs with angular separations between $\theta$ and $\theta + \delta \theta$. The ACF estimator proposed by \citet{1993ApJ...412...64L} is applied, i.e.
\begin{equation}
\omega_\sub{obs}(\theta) = \frac{DD(\theta) - 2DR(\theta) + RR(\theta)}{RR(\theta)},
\end{equation}
where $DD(\theta)$, $DR(\theta)$ and $RR(\theta)$ are the probability of finding galaxy-galaxy, galaxy-random and random-random pairs respectively. These probabilities are calculated by counting all pairs at separations between $\theta$ to $\theta + \delta \theta$, and normalising by the total number of pairs. Estimates of $DR(\theta)$ and $RR(\theta)$ require a catalogue of uniformly-distributed random points. This is generated by a random Poisson process. For each field, the random catalogue contains 10,000 points uniformly placed within the corresponding survey regions. We measure the ACFs in logarithmic bins, and estimate errors by bootstrap resampling \citep{1986MNRAS.223P..21L}. We construct bootstrap subsamples by replacing individual galaxies and perform the resampling for $\sim 500$ times. This approach is also used in \citet{2014ApJ...793...17B} and \citet{2016ApJ...821..123H}. In order to investigate the clustering dependence on both stellar mass and UV magnitude, the total sample is split into subsamples. We choose bins such that they satisfy the $M_\star - M_\sub{UV}$ relation at each redshift. The bin cuts are listed in Table \ref{sm1}.
\par
Since the area of each survey region is finite, the observed ACFs are affected by border effects. This is corrected by an additive constant, which is known as the intergral constrain (IC). Following \citet{1999MNRAS.307..703R}, we have
\begin{equation}
\omega_\sub{true}(\theta) = \omega_\sub{obs}(\theta) + \rm IC,
\end{equation}
with
\begin{equation}
\textrm{IC} = \dfrac{1}{\Omega^2} \iint \omega_\sub{true}(\theta) \, d\Omega_1 d\Omega_2 = \dfrac{\sum_i RR(\theta_i) \omega_\sub{true}(\theta_i) }{\sum_i RR(\theta_i)},
\end{equation}
where $\omega_\sub{true}(\theta)$ is the fitting model of the ACF.
\par
We assume the ACFs to be a power law
\begin{equation}
\omega_\sub{true}(\theta) = A_\omega \left( \frac{\theta}{1''} \right)^{-\beta},
\end{equation}
and construct the log-likelihood using
\begin{equation} \label{likelihood}
\ln \mathcal{L} = -\frac{1}{2} \sum\limits_\sub{fields} \sum\limits_i \left[ \frac{\omega_\sub{obs}(\theta_i) - A_\omega(\theta_i^{-\beta}-\textrm{IC}/A_\omega)}{\sigma(\theta_i)}  \right]^2,
\end{equation}
where sums are over all bins $i$ and over all survey fields. This is equivalent to measuring the average ACF of all fields using inverse-variance weighting. Since this approach requires an estimate of the ACF in each individual field, we only include fields that are deep enough such that the mean separation of galaxy pairs is smaller than $100$ arcsec. The number of LBGs that enter into the analysis for each subsample is given in Table \ref{sm1}. Moreover, the dependence of the IC on the fitting model results in some degeneracy between $A_\omega$ and $\beta$ \citep{2006ApJ...642...63L}, we therefore fix $\beta = 0.6$ following \citet{2006ApJ...642...63L} and \citet{2014ApJ...793...17B}.  In addition, since the area of XDF, HUDF-091, and HUDF-092 is only 4.7 $\textrm{arcmin}^2$ (i.e. one WFC3/IR pointing), the counted number of galaxy pairs in these fields decreases when the angular separation is greater than $\sim 140$ arcsec. We therefore only include separations smaller than that in the likelihood function. 
\par
The amplitude of the ACF $A_\omega$ could be weakened by contamination of lower redshift sources. We reduce this effect by removing all LBGs whose best-fit photometric redshift indicates that it might be a low redshift contaminant ($z_\sub{phot}<2$). It is also noted that \citet{2016ApJ...821..123H} used the contamination fraction estimated by \citet{2015ApJ...803...34B} to correct for this effect. They found that the difference is insignificant compared with the statistical error. Thus, no further treatment is employed to correct the effect of contamination.
\par

\begin{figure*}
  \centerline{\includegraphics[width=17cm]{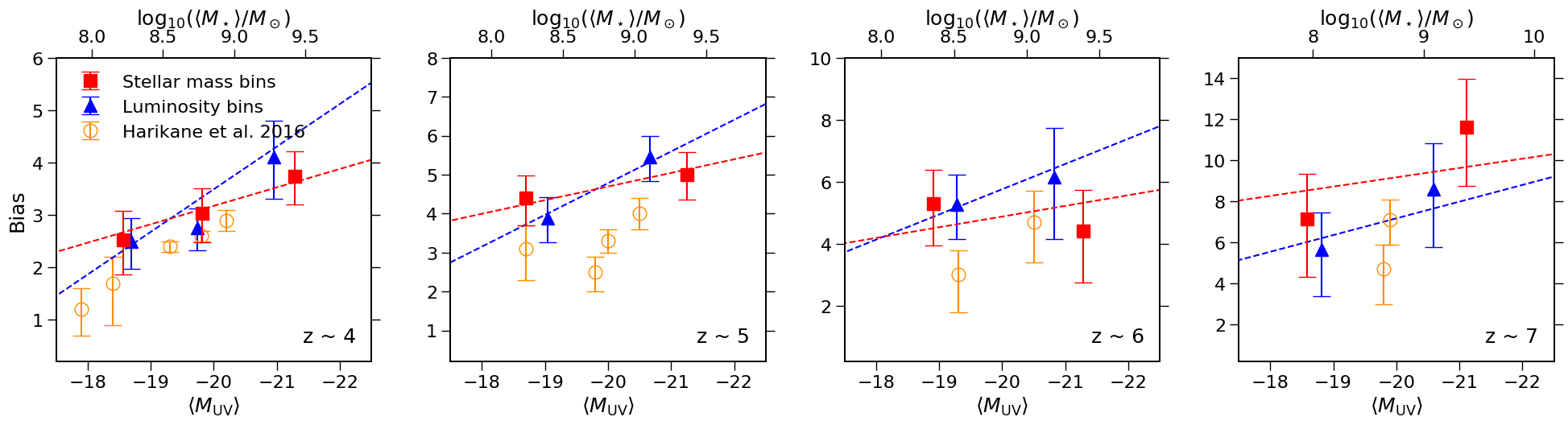}}
  \caption{Measured galaxy biases as a function of mean stellar mass and mean UV flux. The bias is computed by equation \ref{eqn_bias1} and \ref{eqn_bias2}. For all panels, squares (triangles) are corresponding to stellar mass (luminosity) split samples, and are plotted against the top (bottom) axis. Dashed lines are best-fit linear relations of the biases, which are the same with those in Figure \ref{bias2}. Scales of these double x-axes are such chosen that they satisfy the $M_\star - M_\sub{UV}$ relation at each redshift. Orange circles with error bars are biases estimated by \citet{2016ApJ...821..123H} as a function of mean UV magnitude, which are listed in their Table 5.}
  \label{bias1}
\end{figure*}

\subsection{Estimating the correlation length and bias} 
Real space parameters are obtained by applying the Limber transform to the ACFs. The real-space correlation $\xi(r)$ provides three dimensional information on galaxy clustering, which is also approximated by a power law,
\begin{equation}
\xi(r) = \left( \frac{r}{r_0} \right)^{-\gamma},
\end{equation}
 where $r_0$ is called the correlation length. In this case, the Limber transform takes the form \citep{1980lssu.book.....P}
\begin{equation}
\beta = \gamma - 1,
\end{equation} 
and
\begin{equation}
A_\omega = r_0^{\gamma} B \left(\frac{1}{2}, \frac{\gamma - 1}{2} \right) \frac{\int^\infty_0 dz \, N^2(z) {d_\sub{H}}^{-1} d_\sub{A}^{1-\gamma}(1+z)^{1-\gamma} }{ \left[\int^\infty_0 dz \, N(z) \right]^2 },
\end{equation}
with
\begin{equation*}
d_\sub{H} = \frac{c}{H(z)}, \quad d_\sub{A} = \frac{1}{1+z} \int^\infty_0 d_\sub{H} \, dz,
\end{equation*} 
where $B(x,y)$ is the beta function, $N(z)$ is the redshift distribution function of sample galaxies, and $H(z)$ is the Hubble parameter as a function of redshift. The above equations link $A_\omega$ and $\beta$ to the power law parameters in the real space. $N(z)$ is estimated using the photometric redshifts of each LBG.
\par
We derive the bias using the ratio between the variance of the galaxy and the matter correlation functions smoothed by a top-hat with radius $8 h^{-1} \rm Mpc$:
\begin{equation} \label{eqn_bias1}
b = \frac{\sigma_\sub{8, \, g}}{\sigma_8}, 
\end{equation} 
where \citep{1980lssu.book.....P}
\begin{equation} \label{eqn_bias2}
\sigma_\sub{8, \, g}^2 = \frac{72(r_0/8h^{-1} \textrm{Mpc})^\gamma}{(3 - \gamma)(4 - \gamma)( 6 - \gamma)2^\gamma}.
\end{equation}
\par

\begin{figure*}
  \centerline{\includegraphics[width=10cm]{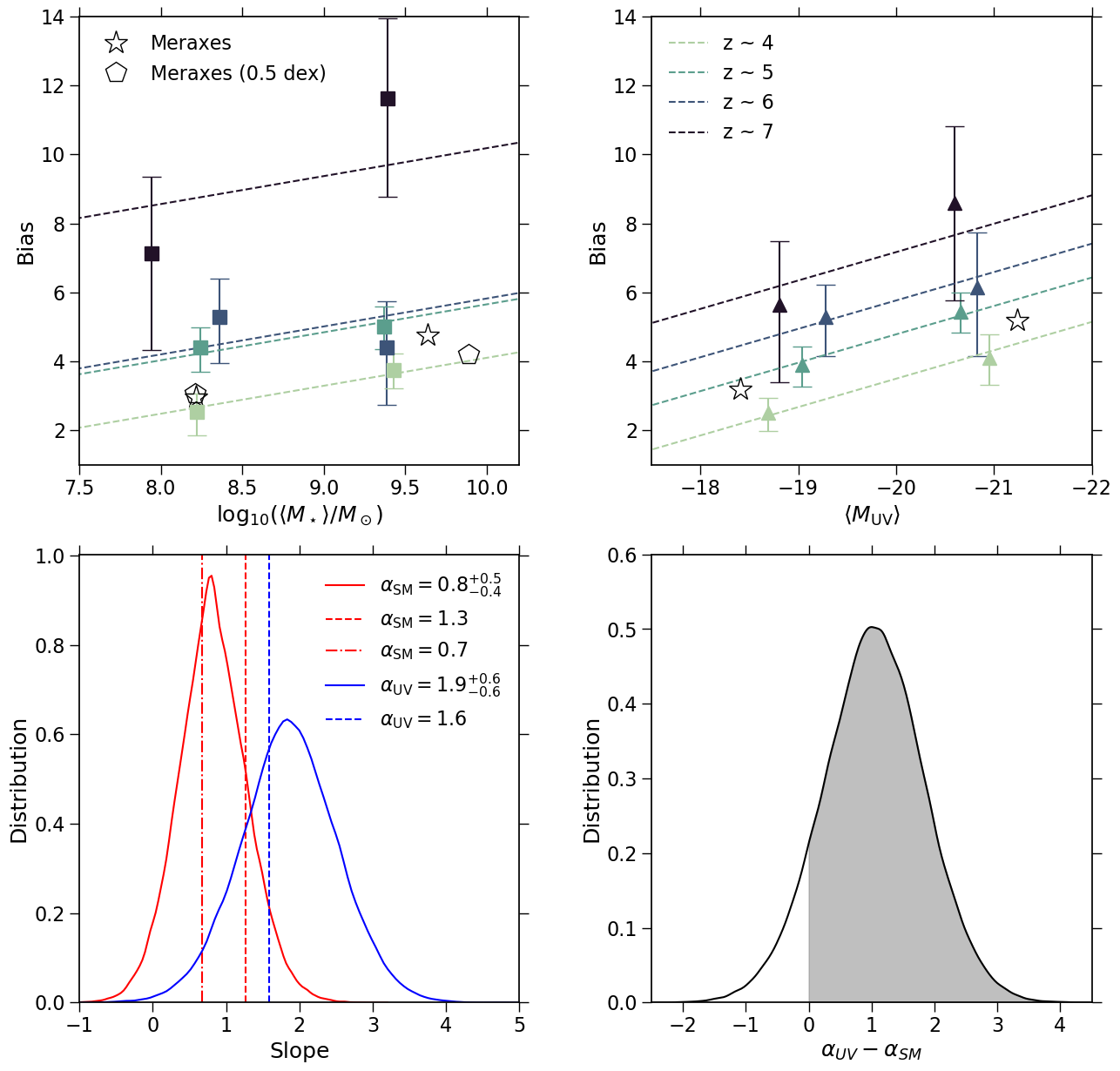}}
  \caption{Results of a straight line fit of measured biases over all redshifts (assuming the same slope). In top panels, dashed lines are the best-fit results. Squares and triangles with error bars are the data that is used for the fits. Empty stars and pentagons are predicted biases from the SAM \textsc{Meraxes}. When computing those pentagons, 0.5 dex Gaussian scatters are added to the stellar mass of each model galaxy. In the bottom left panel, solid lines depict the marginalised distributions of the slope. We show the medians and $1-\sigma$ percentiles of the distributions in the top right corner. Dashed vertical lines show the slopes derived from the model, and the dot dashed line gives the result in the case where the model stellar mass has scatter added. The values of these vertical lines are also shown in the top right corner. In this panel, all red and blue lines correspond to stellar mass and luminosity split samples respectively. Bottom right panel shows the distribution of $\alpha_\sub{UV} - \alpha_\sub{SM}$ obtained by subtracting the samples of $\alpha_\sub{UV}$ and $\alpha_\sub{SM}$. The area of the shaded region is $\simeq 90 \%$, which is the probability that $\alpha_\sub{UV} > \alpha_\sub{SM}$.}
  \label{bias2}
\end{figure*}

\begin{table*}
  \centering
  \caption{Summary of clustering measurements at $z \sim 4 - 7$.}
  \label{sm1}
  \renewcommand{\arraystretch}{1.2}
  \begin{tabular}{c|c|c|c|c|c|c|c|c}
    \hline
    \hline
     $\bar{z}$ & Sample & Cut & Number & $\overline{M}_\star$ & $\overline{M}_\sub{UV}$  & $A_\omega$ & $r_0$ & $b$  \\
     (1) & (2) & (3) & (4) & (5) & (6) & (7) & (8) & (9)\\
    \hline
\multirow{6}{*}{3.8} &\multirow{3}{*}{Stellar mass} &$10^{7.86} M_\odot < M_* < 10^{8.49} M_\odot$ & 1636 & $10^{8.22}$ & $-19.28$ & $0.30^{+0.18}_{-0.16}$ & $2.7^{+0.8}_{-0.9}$ & $2.5^{+0.5}_{-0.7}$ \\ \cline{3-9}                                                                                                                                                                                                                  
 & &$10^{8.49} M_\odot < M_* < 10^{9.01} M_\odot$ & 1691 & $10^{8.77}$ & $-19.82$ & $0.46^{+0.18}_{-0.18}$ & $3.4^{+0.7}_{-0.8}$ & $3.0^{+0.5}_{-0.6}$ \\ \cline{3-9}                                               
 & &$M_* > 10^{9.01} M_\odot$ & 1514 & $10^{9.43}$ & $-20.44$ & $0.70^{+0.20}_{-0.20}$ & $4.4^{+0.7}_{-0.8}$ & $3.7^{+0.5}_{-0.5}$ \\ \cline{2-3} \cline{3-9}                                                       
 &\multirow{3}{*}{Luminosity} &$-19.16 < M_{\rm UV} < -17.74$ & 2012 & $10^{8.71}$ & $-18.70$ & $0.30^{+0.12}_{-0.14}$ & $2.8^{+0.6}_{-0.7}$ & $2.5^{+0.4}_{-0.5}$ \\ \cline{3-9}                                   
 & &$-20.34 < M_{\rm UV} < -19.16$ & 2460 & $10^{9.06}$ & $-19.74$ & $0.40^{+0.12}_{-0.12}$ & $3.0^{+0.5}_{-0.6}$ & $2.8^{+0.4}_{-0.4}$ \\ \cline{3-9}                                                              
 & &$M_{\rm UV} < -20.34$ & 957 & $10^{9.52}$ & $-20.95$ & $0.82^{+0.32}_{-0.32}$ & $4.9^{+1.0}_{-1.2}$ & $4.1^{+0.7}_{-0.8}$ \\ \cline{2-3} \cline{1-9}                                                            
\multirow{4}{*}{5.0} &\multirow{2}{*}{Stellar mass} &$M_* < 10^{8.65} M_\odot$ & 1191 & $10^{8.24}$ & $-19.62$ & $0.82^{+0.26}_{-0.26}$ & $4.3^{+0.7}_{-0.8}$ & $4.4^{+0.6}_{-0.7}$ \\ \cline{3-9}                    
 & &$M_* > 10^{8.65} M_\odot$ & 1554 & $10^{9.37}$ & $-20.56$ & $1.08^{+0.26}_{-0.28}$ & $4.9^{+0.7}_{-0.8}$ & $5.0^{+0.6}_{-0.7}$ \\ \cline{2-3} \cline{3-9}                                                       
 &\multirow{2}{*}{Luminosity} &$M_{\rm UV} > -19.62$ & 1007 & $10^{9.04}$ & $-19.04$ & $0.72^{+0.22}_{-0.24}$ & $3.7^{+0.7}_{-0.7}$ & $3.9^{+0.5}_{-0.6}$ \\ \cline{3-9}                                            
 & &$M_{\rm UV} < -19.62$ & 1661 & $10^{9.47}$ & $-20.66$ & $1.30^{+0.28}_{-0.30}$ & $5.4^{+0.7}_{-0.7}$ & $5.4^{+0.5}_{-0.6}$ \\ \cline{2-3} \cline{1-9}                                                           
\multirow{4}{*}{5.9} &\multirow{2}{*}{Stellar mass} &$M_* < 10^{8.83} M_\odot$ & 342 & $10^{8.36}$ & $-19.80$ & $1.00^{+0.50}_{-0.50}$ & $4.5^{+1.2}_{-1.4}$ & $5.3^{+1.1}_{-1.3}$ \\ \cline{3-9}                   
 & &$M_* > 10^{8.83} M_\odot$ & 293 & $10^{9.39}$ & $-20.46$ & $0.76^{+0.66}_{-0.66}$ & $3.5^{+1.4}_{-1.6}$ & $4.4^{+1.3}_{-1.7}$ \\ \cline{2-3} \cline{3-9}                                                        
 &\multirow{2}{*}{Luminosity} &$M_{\rm UV} > -19.99$ & 450 & $10^{9.38}$ & $-19.28$ & $1.12^{+0.48}_{-0.46}$ & $4.5^{+1.0}_{-1.2}$ & $5.3^{+0.9}_{-1.1}$ \\ \cline{3-9}                                             
 & &$M_{\rm UV} < -19.99$ & 205 & $10^{9.43}$ & $-20.83$ & $1.74^{+1.16}_{-1.14}$ & $5.3^{+1.8}_{-2.1}$ & $6.1^{+1.6}_{-2.0}$ \\ \cline{2-3} \cline{1-9}                                                            
\multirow{4}{*}{6.8} &\multirow{2}{*}{Stellar mass} &$M_* < 10^{8.41} M_\odot$ & 118 & $10^{7.94}$ & $-19.62$ & $0.84^{+0.80}_{-0.80}$ & $5.4^{+2.3}_{-2.6}$ & $7.1^{+2.2}_{-2.8}$ \\ \cline{3-9}                   
 & &$M_* > 10^{8.41} M_\odot$ & 149 & $10^{9.39}$ & $-20.53$ & $3.90^{+1.84}_{-1.84}$ & $10.3^{+2.7}_{-3.1}$ & $11.6^{+2.3}_{-2.8}$ \\ \cline{2-3} \cline{3-9}                                                      
 &\multirow{2}{*}{Luminosity} &$M_{\rm UV} > -19.40$ & 130 & $10^{9.15}$ & $-18.81$ & $0.52^{+0.54}_{-0.52}$ & $4.1^{+1.8}_{-2.1}$ & $5.7^{+1.8}_{-2.3}$ \\ \cline{3-9}                                             
 & &$M_{\rm UV} < -19.40$ & 200 & $10^{9.56}$ & $-20.60$ & $1.82^{+1.22}_{-1.22}$ & $6.9^{+2.3}_{-2.8}$ & $8.6^{+2.2}_{-2.8}$ \\ \cline{2-3} \cline{3-9}
\hline
  \end{tabular}
  \begin{tablenotes}
    \item Notes. - (1) Mean redshift. (2) Sample name. (3) Stellar mass and luminosity cuts. The unit of stellar mass cut is $M_\odot$. (4) Number of galaxies in the subsample. (5) Mean stellar mass in a unit of $M_\odot$. (6) Mean UV fluxes. (7) Amplitude of ACF with fixed $\beta = 0.6$. (8) Correlation length in a unit of $h^{-1}$Mpc. (9) Bias.
  \end{tablenotes}
  \label{res}
\end{table*}

\section{Results and discussion} \label{results}
Figure \ref{acfs} shows measured ACFs and best-fit power laws for both stellar mass and luminosity subsamples. Results for the angular correlation function amplitude ($A_\omega$), the correlation length ($r_0$), and bias ($b$) are summarised in Table \ref{res}. All quantities given in the table are the most probable value and the corresponding $1\sigma$ error.
\par
Clustering segregation is observed with both stellar mass and luminosity. From Figure \ref{acfs}, it is clear that the ACFs increase with luminosity at all redshifts. Similar trends are also observed for different stellar mass subsamples. However, the segregation with stellar mass is found to be weaker than with luminosity. To further compare the clustering dependence of these two properties, we plot the measured bias as a function of mean stellar mass and flux in Figure \ref{bias1}. The bias increases with stellar mass, from $b = 2.7^{+0.5}_{-0.6}$ to $b = 3.6^{+0.5}_{-0.5}$ at $z \sim 4$, and $b = 4.2^{+0.6}_{-0.7}$ to $4.8^{+0.6}_{-0.7}$ at $z \sim 5$. The segregation of bias with luminosity is more obvious. At $z \sim 4$, the bias for the brightest sample is $b = 4.1^{+0.7}_{-0.8}$, while that of the faintest sample is smaller at $b = 2.5^{+0.4}_{-0.5}$. Similarly, at $z \sim 5$, the bias has an increase between the faint and bright samples, from $b = 3.8^{+0.5}_{-0.6}$ to $b = 5.3^{+0.6}_{-0.6}$. At $z \gtrsim 6$, the LBG sample is smaller, and uncertainties become much larger. While there is a small increase of the bias with mean UV flux at $\bar{z} = 5.9$, no significant segregation with stellar mass is detected. For samples at $z \sim 7$, we find that the bias increases from $b = 7.4^{+2.0}_{-2.6}$ to $b = 11.7^{+2.2}_{-2.7}$, and from $b = 5.6^{+1.8}_{-2.2}$ to $b = 8.6^{+2.2}_{-2.8}$, for stellar mass and luminosity bins respectively. As a comparison, we also plot the bias estimated by \citet{2016ApJ...821..123H} from their power law fits as a function of mean UV magnitude. We find that the trends of clustering dependence on luminosity are consistent, while the offsets on biases themselves could be due to different methods of computing them. To summarise, we find that both more massive and more luminous galaxies are more highly clustered, implying that they are hosted by more massive dark matter halos.
\par
On the other hand, the comparison of segregation between stellar mass and luminosity disagrees with our prior expectations, especially at $\bar{z} = 3.8$ and $5.0$. In particular, we find that the clustering dependence is larger for luminosity than for stellar mass.  In order to quantify this trend, we fit the measured biases of the lightest (faintest) and heaviest (brightest) by straight lines, i.e.
\begin{equation}
b = \alpha_\sub{SM} \log_{10}(M_\star / M_\odot) + \rm const,
\end{equation}
and
\begin{equation} \label{aUV}
b = -1.1 \, \alpha_\sub{UV} \log_{10}(L_\sub{UV} / L_\odot) + \rm const.
\end{equation}
If clustering segregation with both properties were the same, one would expect that the ratio of the slopes should recover the slope of the $M_\star - M_\sub{UV}$ relation. Therefore, we include a correction factor of $-1.1$ to equation \ref{aUV} in order to make $\alpha_\sub{SM}$ and $\alpha_\sub{UV}$ comparable. The value $-1.1$ is based on the results in Table \ref{LTM}. If $M_\star \propto L$, this factor would become unity. We combine the measured biases over all redshifts, assume the same slope but different intercepts for each redshift, and fit these five parameters using the \textsc{emcee} MCMC sampler developed by \citet{2013PASP..125..306F}, assuming flat priors. Best-fit results are shown in the left and middle panels of Figure \ref{bias2}. We illustrate the marginalised distributions of the slopes for stellar mass and luminosity subsamples as solid red and blue lines respectively in the bottom left panel of Figure \ref{bias2}. The median and $1-\sigma$ percentiles of the slopes are summarised in the top right corner. We conclude that the slope in the stellar mass case is systematically smaller than in the luminosity case, which indicates larger clustering segregation with luminosity. The significance of this trend can be seen from the bottom right panel of the figure. The probability that $\alpha_\sub{UV} > \alpha_\sub{SM}$ is $\simeq 90 \%$ (shaded region). 
\par
However, UV magnitude only probes recent starbursts of a galaxy, while stellar mass is an integrated quantity, which reflects the whole star formation history. This argument suggests that a tighter relation is expected between stellar and halo mass, and therefore, clustering segregation with stellar mass should be larger. Both observational effects and astrophysical reasons could be responsible for this discrepancy. We investigate this issue further by comparing the results at $z \sim 4$ with predictions from the SAM \textsc{Meraxes} in the next section. 

\section{Comparison with Meraxes} \label{compare}
\subsection{Overview of the model}
The \textsc{Meraxes} SAM is part of the Dark-ages, Reionization And Galaxy-formation Observables Numerical Simulation (DRAGONS)\footnote[1]{\label{fn2}http://dragons.ph.unimelb.edu.au/} project. It is designed to self-consistently model galaxy formation and evolution during and after the epoch of the reionisation. Physics implemented in the model is described in \citet{2016MNRAS.462..250M} and \citet{2017MNRAS.472.2009Q}. In the present work, the model is run on the extended \textit{Tiamat} N-body simulation \citep{2016MNRAS.459.3025P,2017MNRAS.472.3659P}. The simulation has $2160^3$ dark matter particles, with particle mass $m_\sub{p} = 2.64 \times 10^6 h^{-1}M_\odot$, and side length $67.8 \, h^{-1}$Mpc. The simulation outputs 101 snapshots between $z = 35$ and $z = 5$ with time steps separated by $\sim 11$ Myr, and additional 63 snapshots from $z = 5$ to $z = 1.8$. The time steps of these additional snapshots are evenly spaced in dynamical time. We adopt the same parameter configuration as in \citet{2017MNRAS.472.2009Q} for the SAM. All model stellar mass is subtracted by 0.24 dex in order to convert from a \citet{1955ApJ...121..161S} IMF to a \citet{2003PASP..115..763C} IMF.
\par
The computation of photometry of model galaxies is described in \citet{2016MNRAS.462..235L}. Progenitors of each model galaxy are treated a series of single stellar populations (SSPs), and integrated over the whole star formation history. We assume constant metallicity $Z = 0.001$ for the SSP. Recent hydrodynamical simulations \citep{2016MNRAS.456.2140M,2016MNRAS.462.3265D} predict a mass-metallicity relation that is close to this value at $z \sim 4$. \cite{2017MNRAS.470.3006C} point out that intrinsic luminosities only weakly depend on the metallicity. Therefore, this assumption is valid for this work. SSPs templates are generated by \textsc{starburst99} \cite{1999ApJS..123....3L,2005ApJ...621..695V,2010ApJS..189..309L,2014ApJS..212...14L} assuming the same IMF with \textsc{meraxes}. Modelling of the Lyman absorption due to the intergalactic medium (IGM) is critical for the simulated LBG selection. We update the transmission curve of the IGM using a recent study from \citet{2014MNRAS.442.1805I}, which predicts a more consistent redshift distribution for selected model LBGs. In addition, we also update the dust correction using an empirical model proposed by \citet{2015ApJ...813...21M}, which is applicable at $z < 4$.
\par
\begin{figure}
  \hfill\includegraphics[width=7cm]{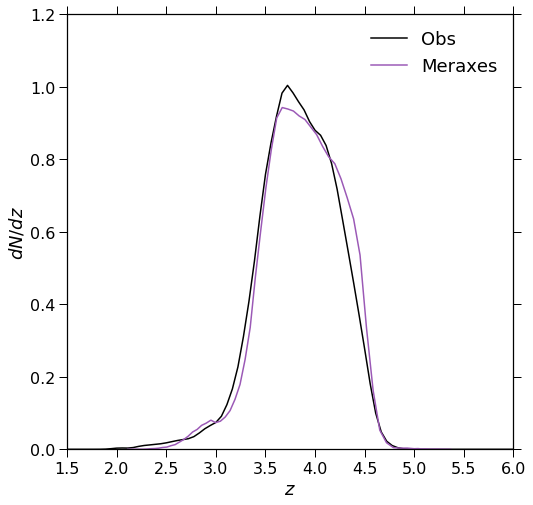}\hspace*{\fill}
  \caption{Example redshift distribution of observed and selected model galaxies in the $z \sim 4$ LBG sample. The observed distribution is estimated by photometric redshifts. Our color selection on the model galaxies results in a very well-matched redshift distribution. The field depth of the XDF is used to select model galaxies.}
  \label{distri}
\end{figure}

 
We follow \citet{2016MNRAS.461..176P,2017MNRAS.472.1995P} to select model LBGs and calculate the ACFs. This approach mimics the incompleteness of the LBG sample by adding photometric scatter to the magnitudes of each LBG selection band. The level of the scatter is given by the $1\sigma$ field detection limits. In the present work, we assemble model LBGs according to the flux limits of the deepest field in the our observations, i.e. the XDF. The resulting redshift distribution of model selected LBGs is shown in Figure \ref{distri}, which agrees with the observed one estimated by photometric redshifts. In terms of the determination of the ACF, we compute the real-space correlation function across a sequence of snapshots directly from the spatial coordinates of each galaxy, and convert to an ACF by the Limber transform. The readers are referred to \citet{2016MNRAS.461..176P,2017MNRAS.472.1995P} for a more detailed description.

\begin{figure}
  \includegraphics[width=8cm]{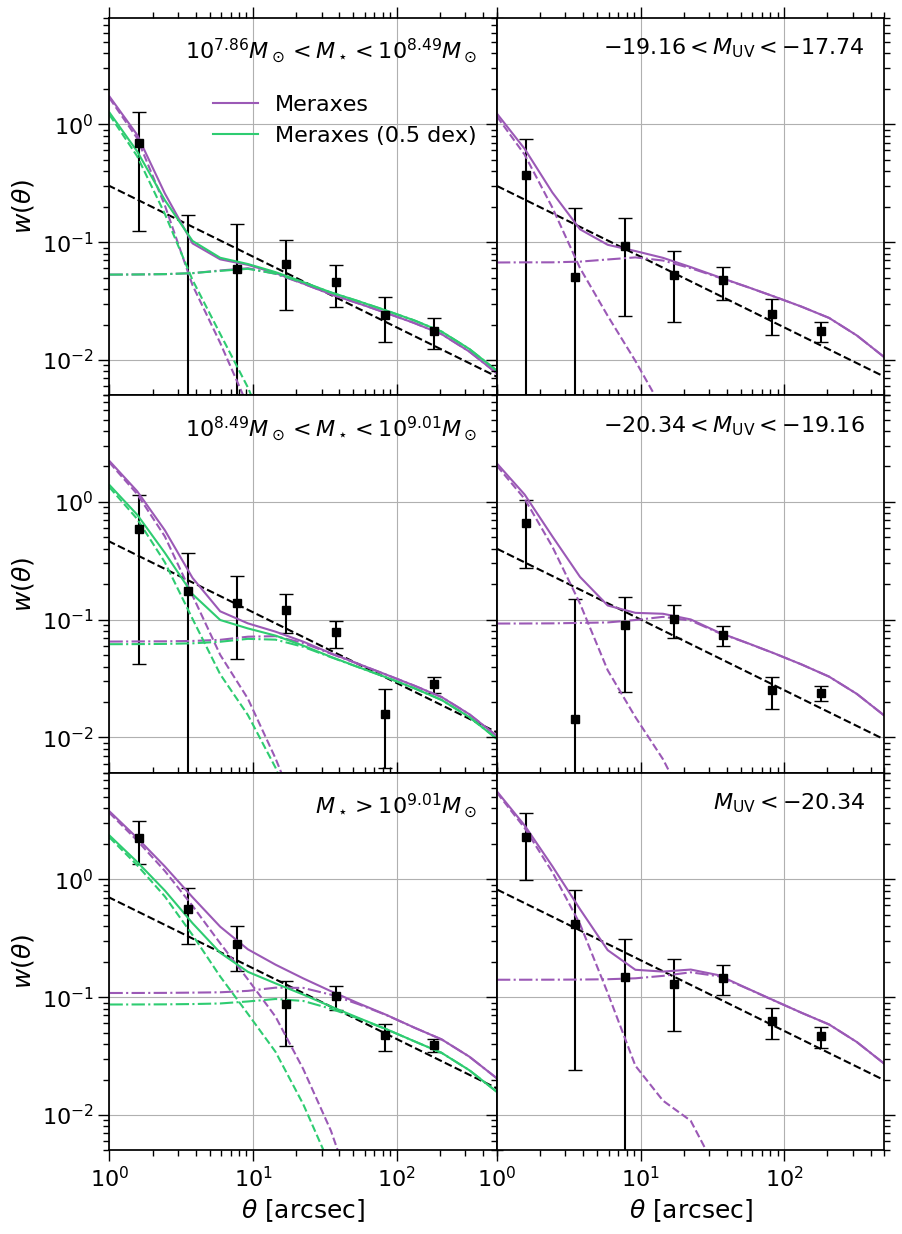} 
  \caption{Comparison between observed and model predicted ACFs at $z \sim 4$. Black squares with error bars and black dash lines are observed ACFs and the corresponding best-fit power laws, which are the same with those in the top left panel of Figure \ref{acfs}. Purple lines are ACFs estimated from Meraxes. Purple dashed and dot dashed lines are the corresponding 1-halo and 2-halo terms, which is calculated by counting galaxy pairs in the same and different FoF groups. Green lines represent the model predictions from the sample where 0.5 dex Gaussian scatters are added to stellar mass.}
  \label{acf2}
\end{figure}

\subsection{Results} \label{scatter}
We focus on the comparison between our model and observations at $z \sim 4$, where the measurements have the smallest errors. We plot predicted ACFs together with measured ACFs in Figure \ref{acf2}. The model ACFs agree very well with observations, and reproduce the observed clustering dependence on both stellar mass and luminosity. In order to check whether the model predicts larger clustering segregation with luminosity as indicated by our observations, we therefore calculate the bias from the model by $b^2 = \xi(r)/\xi_\sub{DM}(r, z)$, where $\xi(r)$ and $\xi_\sub{DM}(r, z)$ are the real-space correlation functions of galaxies and dark matter. An average value is taken in the range $5 \, \textrm{Mpc} \leq r \leq 10 \, \textrm{Mpc}$. The estimated biases for the lightest (faintest) and heaviest (brightest) samples are shown using star symbols in the left (middle) panel in Figure \ref{bias2}. Subsequently, we derive the slope for these two points for comparison with observational data in Figure \ref{bias2}. 
We conclude that the measured variation of bias with stellar mass and luminosity is consistent with predictions and that the clustering segregation with luminosity is larger than stellar mass in the model. In other words, the model predictions also contradict our expectation that stellar mass should be more tightly correlated with halo mass. The physical interpretation behind this could be complex, and we defer it to a subsequent paper.
\par
The observed clustering dependence on stellar mass could be weakened due to observational uncertainties on the estimations of stellar mass. The uncertainties, for instance, can be due to the low S/N ratio of Spitzer data. We demonstrate this effect by adding 0.5 dex of Gaussian scatter to the stellar mass of selected model LBGs and remeasure the clustering in stellar mass bins. This level of scatter is also used in abundance matching studies \citep[e.g.][]{2013MNRAS.428.3121M,2013ApJ...770...57B}. The recalculated ACFs are shown as green lines in Figure \ref{acf2}. For the most massive bin, the ACF decreases at all scales, while for the other two bins, scatter in stellar mass only affects the small scale correlation functions. We also recalculate the bias and the corresponding slope. The results are demonstrated in Figure \ref{bias2}. In the right panel of Figure \ref{bias2}, the dot dashed line represents the slope in the case where scatter is added to the stellar mass, and shows that scatter reduces the slope relative to that of the original model, from $\alpha_\sub{SM} = 1.3$ to $\alpha_\sub{SM} = 0.7$. This effect is most significant for the most massive bins, since the massive end has fewer galaxies. We can use this systematic error of model $\alpha_\sub{SM}$ introduced by adding scatter to the stellar mass to estimate the effect on our measured $\alpha_\sub{SM}$. This indicates that accounting for uncertainties in stellar mass leads to clustering segregation that is similar for both mass and UV-luminosity, but which is not larger with stellar mass. Hence, although uncertainties in stellar mass can weaken the clustering dependence, the unexpected trend could still result from physical reasons.

\subsubsection{Satellite galaxies}
Another interesting finding from the comparison between observations and the model is the deviation of the ACFs from a power law at small scales ($\theta < 10$ arcsec). In the model, this is due to the 1-halo term, arising from multiple halo occupation, where more than one galaxy resides in the same halo. To provide evidence of this multiple halo occupation, we calculate the 1-halo and 2-halo terms explicitly from \textsc{Meraxes} by counting galaxy pairs in the same and different FoF groups. These two terms are shown as dashed and dot dashed lines in Figure \ref{acf2}, respectively, demonstrating that the steep increase of the ACFs at small scales is due to multiple halo occupation. Consistency is also found between observations and the model. It can also be seen that the transition of the model ACFs between the 1-halo and 2-halo terms becomes more rapid with decreasing stellar mass. However, there is no such trend in luminosity. This finding may suggest an additional feature of clustering segregation with stellar mass and luminosity, implying different satellite properties for the two cases. However, we caution that the satellite properties of \textsc{Meraxes} have yet to be fully explored and compared with observations, and that achieving realistic recent satellite star formation histories has traitionally been a challenging task for SAMs. This difference can also be seen from the observed ACFs but with large uncertainties. At the very bright end, the smooth transition between 1-halo and 2-halo terms is observed in \citet{2018PASJ...70S..11H}, and explained using non-linear bias \citep{2016MNRAS.463..270J}. Larger surveys with more complete samples might be used to investigate this phenomenon in more details for fainter and less massive galaxies.

\section{Summary} \label{sum}
We have carried out a clustering analysis of LBGs over the range $z \sim 4 - 7$, with emphasis on the comparison between clustering segregation with stellar mass and luminosity. We also compare our measurements with predictions from the \textsc{Meraxes} SAM. Our findings can be summarised as follows:
\begin{itemize}
  \item The observed ACF amplitude and bias generally increase with stellar mass and luminosity over $z \sim 4 - 7$. The ACFs obtained from the model are consistent with observations, and reproduce clustering segregation with both stellar mass and luminosity. This suggests that more luminous and massive galaxies are more clustered, and hence hosted by more massive dark matter halos.
  \item By combining measurements over all redshifts, a systematic difference is found between clustering segregation with stellar mass and luminosity. In particular, it is observed that clustering strength is more tightly correlated with luminosity. This is in contrast to the expectation that stellar mass should be more tightly correlated with halo mass since stellar mass reflects the whole star formation history, while UV magnitude only corresponds to recent star formation. We find that the model also predicts this surprising result of larger clustering segregation with luminosity.
  \item At $z \sim 4$, the model predicts that the transition between the 1-halo and 2-halo terms of the ACFs is smoother for larger stellar mass galaxies, and that this trend does not appear for samples split by luminosity. Observations show similar behaviour but with large error bars. This might suggest that samples split by stellar mass and luminosity have quite different satellite properties. 
\end{itemize}
Our results extend to higher redshift findings from the recent study at $z \sim 3$ by \citet{2018A&A...612A..42D}, who carried out a clustering analysis of 3236 galaxies discovered in the VIMOS Ultra Deep Survey. They measured the real space correlation functions, and also found a larger difference in the correlation lengths when splitting the sample by luminosity than by stellar mass. This unexpected trend of clustering segregation with stellar mass and luminosity may provide new clues of galaxy formation in the early universe. For instance, some high-redshift galaxies might be formed by a single starburst. In this case, stellar mass and UV-luminosity become similar indicators of the star formation history. Our study also motivates future spectroscopic surveys with the James Webb Space Telescope (JWST), which will provide more complete samples and more accurate stellar mass measurements substantially reducing the systematic errors in clustering studies.

\section*{Acknowledgement}
Parts of this work were performed on the gSTAR national facility at Swinburne University of Technology. gSTAR is funded by Swinburne and the Australian Government's Education Investment Fund. This research was also supported by the Australian Research Council Centre of Excellence for All Sky Astrophysics in 3 Dimensions (ASTRO 3D), through project number CE170100013.

\bibliographystyle{\dir mn2e}
\bibliography{reference}
\bsp
\label{lastpage}
\end{document}